\documentclass[aps,preprint,10pt,twocolumn]{revtex4}%
\usepackage{amsfonts}
\usepackage{amsmath}
\usepackage{amssymb}
\usepackage{graphicx}%
\begin{document}
\preprint{ }
\title{Quantum Spin Effect and Short-Range Order above the Curie Temperature}
\author{R. Y. Gu and V. P. Antropov}
\affiliation{Condensed Matter Physics, Ames Laboratory, Ames, IA 50011}
\date{\today}

\begin{abstract}
Using quantum Heisenberg model calculations with Green's function technique
generalized for arbitrary spins, we found that for a system of small spins the
quantum spin effects significantly contribute to the magnetic short-range
order and strongly affect physical properties of magnets. The spin dynamics
investigation confirms that these quantum spin effects favor the persistence
of propagating spin-wave excitations above the Curie temperature. Our
investigation suggests a reconsideration of prevailing point of view on finite
temperature magnetism to include quantum effects and the magnetic short-range order.

\end{abstract}
\maketitle

It is a long-standing debate over the nature of the paramagnetic (PM) state of
the ferromagnetic (FM) materials, particularly in the transition metals. Early
inelastic neutron experiments\cite{mook,lynn,mook2} determined the persistence
of spin-wave (SW) like modes above the Curie temperature ($T_{C}$) in both Ni
and Fe, and these modes were interpreted as the evidence of considerable
magnetic short-range order (MSRO) in the PM\ state\cite{korenman}. The
presence of MSRO was later supported by the spin and angle-resolved
photoemission studies \cite{maetz,haines}. This is rather unusual because
majority of magnetism theories are based on the absence of SW excitations
above $T_{C}$. Moreover, by applying the spherical model (SM) approximation to
the Heisenberg model (HM), Shastry \cite{shastry} \textit{et al.} concluded
that in contrast with the experiment \cite{lynn}, this model, with fairly
long-ranged interactions, has little MSRO and no SW peaks above $T_{C}$ in Fe.
A Similar conclusion was reached by Monte Carlo (MC) spin dynamics simulation
of the classical HM for the same system\cite{shastry2}. Very sophisticated
technique\cite{DMFT} also could not detect any traces of strong MSRO in Fe or
Ni.\ Recent theoretical spin dynamics studies\cite{MSROLDA}, however, have
demonstrated that a strong MSRO is a 'must have' property of the itinerant
magnets while such excitations like SW exist above $T_{C}$ in both localized
and itinerant magnets.

It is puzzling that the applications of HM failed to predict the expected MSRO
because BCC Fe has rather good local moments \cite{hasegawa,hubbard} and to a
large extent HM should be valid \cite{edwards}. What usually is omitted in the
classical MC simulation is the quantum nature of spin. In quantum SM $T_{C}$,
\ the spin correlations and the susceptibility are proportional to $S(S+1),$
while the classical coefficient scales as $S^{2}$; so the quantum effect
contributes a factor $Q_{S}=1+S^{-1}$. For small $S$ this $Q_{S}$ can be
unreasonably large. For example, for $S=1/2$ $Q_{S}=3$ which leads to the
unphysical correlation between nearest-neighbor (NN) spins $\langle
\boldsymbol{S}_{i}\cdot\boldsymbol{S}_{j}\rangle/S^{2}>1$ in the case of NN
coupling of the simple cubic (SC) structure. The same problem appeared in
Ref.\cite{shastry} where in the case of strong MSRO and $S=1$ $\langle
\boldsymbol{S}_{i}\cdot\boldsymbol{S}_{j}\rangle/S^{2}=1.64$. To avoid such
difficulties a pragmatical approach is to scale the relevant quantities by
$S(S+1)$, as was done in Ref. \cite{shastry}. Then, in SM the scaled
quantities are independent of $S,$ so the quantitative results for MSRO and
region of SW existence for $S=1$ and $S=\infty$ are the same\cite{shastry}.
Below we will demonstrate that while the classical HM can describe some degree
of MSRO above $T_{C},$ QSE properly included strongly increases MSRO and
affects its influence on other physical properties.

For the HM hamiltonian $H=-\sum_{(ij)}J_{ij}\boldsymbol{S}_{i}\cdot
\boldsymbol{S}_{j}$ in the PM$\ $state, we use the second-order Green's
function (GF) technique\cite{kondo,shimahara,barabanov,winterfeldt}. To
calculate GF $G_{ij}^{\omega}=\langle\langle S_{i}^{-};S_{j}^{+}\rangle
\rangle_{\omega}$ one applies twice the equation of motion and then decouples
the high-order GF of forms $\langle\langle S_{\rho}^{z}S_{m}^{z}S_{i}%
^{-};S_{j}^{+}\rangle\rangle_{\omega}$ and $\langle\langle(S_{\rho}^{+}%
S_{l}^{-}-S_{\rho}^{-}S_{l}^{+})S_{i}^{-};S_{j}^{+}\rangle\rangle_{\omega}$.
For $S=1/2$ in one-dimensional system Kondo and Yamaji \ (KY) decoupled them
by using a correction parameter $\alpha$ \cite{kondo}. Here we extend their
method for arbitrary $S$ by introducing the following decoupling scheme (for
$i\neq\rho$ and $\rho\neq l)$,%
\begin{gather}
\langle\langle\{S_{\rho}^{+}S_{l}^{-}-S_{\rho}^{-}S_{l}^{+},S_{i}^{-}%
\};S_{j}^{+}\rangle\rangle_{\omega}\rightarrow(1-\nu_{s}\delta_{il}%
)\widetilde{C}_{i\rho}G_{lj}^{\omega}-\widetilde{C}_{il}G_{\rho j}^{\omega
},\nonumber\\
\langle\langle\{S_{\rho}^{z}S_{m}^{z},S_{i}^{-}\};S_{j}^{+}\rangle
\rangle_{\omega}\rightarrow(1-\nu_{s}\delta_{im})\widetilde{C}_{m\rho}%
^{zz}G_{ij}^{\omega},\text{ }\label{decouple1}%
\end{gather}
where $\widetilde{C}_{m\rho}^{zz}=\alpha_{m\rho}C_{m\rho}^{zz}$ and
$\widetilde{C}_{il}=\alpha_{il}C_{il},$ with $\alpha_{m\rho}=\alpha
(1-\delta_{m\rho})+\delta_{m\rho}$ and $C_{m\rho}^{zz}=\langle S_{\rho}%
^{z}S_{m}^{z}\rangle,$ $C_{il}=\langle S_{l}^{+}S_{i}^{-}\rangle=2\langle
S_{l}^{z}S_{i}^{z}\rangle$ are the spin correlations, $\{A,B\}=(AB+BA)/2$ is
the symmetric product of operators, and $\nu_{s}$ is a $S$-dependent constant
which will be determined later. For $S=1/2$ spin operator identities require
$\nu_{s}=1$, and Eq. (\ref{decouple1}) is reduced to the KY decoupling.

Decoupling the high order GF in the equation of motion with
Eq.(\ref{decouple1}) one can obtain the following expression for the dynamic susceptibility%

\begin{equation}
\chi^{+-}(\boldsymbol{q},\omega)=-\frac{2\sum_{n}z_{n}J_{n}C_{n}(1-\gamma
_{n}^{\boldsymbol{q}})}{\omega^{2}-\omega_{\boldsymbol{q}}^{2}}, \label{kaikw}%
\end{equation}
where $n$ is the shell index, $J_{n}$ and $C_{n}$ are $J_{ij}$ and $C_{ij},$
correspondingly. \textit{ }$\gamma_{n}^{\boldsymbol{q}}=z_{n}^{-1}\sum
_{\delta_{n}}(1-e^{i\boldsymbol{q}\cdot\boldsymbol{\delta}_{n}})$ with $z_{n}$
being the total number of sites on\ $n-$th\textit{ }shell\textit{ }and
$\boldsymbol{\delta}_{n}$ being sites on that shell.

The SW excitation spectrum is%

\begin{equation}
\omega_{\boldsymbol{q}}=\left\{  \sum_{n}z_{n}J_{n}(1-\gamma_{n}%
^{\boldsymbol{q}})\left[  D_{n}-\nu_{s}J_{n}\widetilde{C}_{n}%
-J^{\boldsymbol{q}}\widetilde{C}_{n}\right]  \right\}  ^{1/2}\label{wk}%
\end{equation}
where $D_{n}=N^{-1}\sum_{\boldsymbol{k}}J^{\boldsymbol{k}}\gamma
_{n}^{\boldsymbol{k}}\widetilde{C}^{\boldsymbol{k}}$ with $J^{\boldsymbol{k}}$
and $\widetilde{C}^{\boldsymbol{k}}$ being the Fourier transforms of
\ $J_{ij}$ and $\widetilde{C}_{ij}$, correspondingly. $\ $At this stage
$\nu_{s}=(2-S)/3S$ \ is obtained by comparing Eq.(\ref{wk}) with the
well-known result $\omega_{\boldsymbol{q}}=(J^{\boldsymbol{0}}%
-J^{\boldsymbol{q}})S$ in the FM spin correlation limit $C_{n}=2S^{2}/3$.

\ From Eq.(\ref{kaikw}) and the spectral theorem, the spin correlation can be
written as%

\begin{equation}
C^{\boldsymbol{q}}=\sum_{n}z_{n}J_{n}(1-\gamma_{n}^{\boldsymbol{q}}%
)\frac{C_{n}}{\omega_{\boldsymbol{q}}}\coth\frac{\omega_{\boldsymbol{q}}}{2T}.
\label{cq}%
\end{equation}
With the requirements $C_{n}=1/N\sum_{\boldsymbol{q}}C^{\boldsymbol{q}}%
\gamma_{n}^{\boldsymbol{q}}$ and $C_{0}=1/N\sum_{\boldsymbol{q}}%
C^{\boldsymbol{q}}=2S(S+1)/3,$ Eqs.(\ref{wk}) and (\ref{cq}) can be solved
self-consistently. $T_{C}$ is determined by $\chi^{-1}=0$ ($\chi=\chi
^{+-}(0,0)/2$ ). To check the validity of our method, in Fig. 1 we compare our
calculated $T_{C}$ $/T_{C}^{\text{MF}}$and $E_{c}/E_{0}$ for the BCC structure
with the accurate results obtained by the high-temperature-expansion (HTE)
methods\cite{rushbrooke,bowers} and the SM results in the NN coupling case.
Here $T_{C}^{\text{MF}}$ is the Curie temperature in the mean field (MF)
approximation, $E_{c}$ and $E_{0}$ are the total energies at $T_{C}$ and zero
temperature, correspondingly. The parameter $E_{c}/E_{0}$ is a proper measure
of MSRO at $T_{C}$, and in the NN coupling case $E_{c}/E_{0}$ is identical to
the average cosine of angles between NN spins. In the MF approximation there
is no MSRO ($E_{c}/E_{0}=0)$ at and above $T_{C}^{\text{MF}}.$ The existence
of MSRO suppresses $T_{C}$ with respect to $T_{C}^{\text{MF}}$. Such
suppression exists also in the SM and is identical for all $S$ in that case.
In more accurate calculations, however, $T_{C}$ is more suppressed at smaller
$S.$

The MSRO parameter $E_{c}/E_{0}$ demonstrates the increase in MSRO for smaller
$S$. Although in the SM increases even faster ($E_{c}/E_{0}\propto Q_{S}$),
this quantity already is not well defined owing to the appearance of
$E_{c}/E_{0}>1$ e.g. for the SC structure for $S=1/2$ and in
Ref.\cite{shastry} for $S=1$. At this stage the scaling should be introduced
which leads to the elimination of real QSE.

Our formalism allows to obtain the following important result for $T_{C}$ for
$S=\infty$
\begin{equation}
T_{C}=\alpha T_{C}^{\text{SM}}=3T_{C}^{\text{MF}}/(2F+1), \label{tc}%
\end{equation}
where $\alpha=3F/(2F+1)$ with $F=N^{-1}\sum_{\boldsymbol{q}}(1-\gamma
_{1}^{\boldsymbol{q}})^{-1}$. This new and transparent expression provides
another immediate and accurate check of applicability of our generalized GF
formalism. For instance, it gives $T_{C}/(J_{1}S^{2})=1.49,2.11,$ and $3.25$
for SC, BCC and FCC structures which are very close to the corresponding HTE
results $1.45,2.06,$ and $3.18$\cite{bowers}. Eq.(\ref{tc}) clearly indicates
the importance of the correction parameter $\alpha$ introduced above in the GF
decoupling. For $S=\infty$ the parameter $E_{c}/E_{0}=1-F^{-1\text{ }}$is the
same as the one obtained in the SM.

The good agreement between our and HTE results indicates the applicability of
this formalism for the case of arbitrary $S$ and NN interaction. We also
studied a Heisenberg hamiltonian corresponding to a realistic material: we
used extended (four NN) interactions in BCC Fe: $J_{2}/J_{1}=0.5221,$
$J_{3}/J_{1}=0.0056,$ and $J_{4}/J_{1}=-0.0879$\cite{antropov}, where
$J_{1}S^{2}=2.44$ mRy. For $S=\infty$ MC simulation gives $T_{C}^{\text{MC}%
}/T_{C}^{\text{MF}}=0.68\sim0.70,$and $E_{C}/E_{0}=0.38\sim0.41.$ In SM
$T_{C}^{\text{SM}}/T_{C}^{\text{MF}}=0.59$ and $E_{C}/E_{0}=0.40Q_{S}$ for all
$S.$ In our formalism, for $S=1/2,1$ and $\infty,$ $T_{C}/T_{C}^{\text{MF}%
}=0.51,0.57$ and $0.67,$ and their $E_{C}/E_{0}=0.79,0.68$ and $0.41,$
correspondingly. At $S=1$, $T_{C}^{\text{MF}}=2334$K is more than twice higher
than the experimental $1040$K of Fe, our calculated $T_{C}$ is suppressed to
the much lower value $1330$K. Comparing with Fig.1, one can see the additional
suppression of $T_{C}$ with $E_{C}/E_{0}$ being considerably larger, thus
indicating stronger MSRO than in the corresponding NN coupling case. However,
the parameters in Ref.\cite{antropov} have been obtained in the
long-wavelength approximation and can only describe a small MSRO in classical
case. The inset of Fig.2 shows directly $\cos\theta_{n}=\langle\boldsymbol{S}%
_{i}\cdot\boldsymbol{S}_{i+\delta_{n}}\rangle/S^{2}$, giving the details of
the QSE enhancement of MSRO between several neighboring spins.

The spin correlation length $\xi$ is often used to describe the strength of
MSRO. Despite the magnitude of $\cos\theta_{n}$, $\xi$ always tends to
infinity when temperature approaches $T_{C}$ from above, so near $T_{C}$ $\xi$
may not be a parameter that properly reflects MSRO. Above $T_{C}$ the
evaluation of $\xi$ in our formalism is straightforward from the
long-wavelength behavior of the spin-correlation function $C^{\boldsymbol{q}%
}\propto1/(q^{2}+\xi^{-2}).$ We found that at fixed $T/T_{C},$ $\xi$ always
increases as $S$ becomes smaller, in contrast to the SM where $\xi$ is
independent of $S$. At $T=1.1T_{C},$ \ $\xi=4.1,3.7$ and $2.8$ for $S=1/2,1$
and $\infty,$ again demonstrating the QSE enhancement of MSRO from another prospective.

Now let us analyze the SW excitations. In the standard magnetism theories such
as the random phase approximation\cite{tahir2} and its various modified
versions\cite{callen}, SW exist due to the magnetic long-range order, so its
spectrum is renormalized to zero at $T_{C}.$ In our formalism SW comes from
the short-range spin correlations and the long-range order is no longer a
prerequisite\ for its existence, so SW spectrum can be finite at $T_{C}.$ In
Fig.2 we plot the calculated SW spectrum obtained from Eq.(\ref{wk}) at
$T_{C}$. To demonstrate the $S$-dependence of the SW renormalisation, the SW
spectrum at $T=0$ (the FM case) is also plotted with all $\omega
_{\boldsymbol{q}}$ scaled by $S$.

Let us estimate the renormalisation factor in the BCC Fe\cite{lynn} where SW
modes have been observed above the middle of the Brillouin zone along the
(110) direction, $\boldsymbol{Q}=(\frac{\pi}{2}\frac{\pi}{2}0)$ (lattice
constant $a=1$). The SW renormalisation factors $\omega_{\boldsymbol{Q}}%
(T_{C})/$ $\omega_{\boldsymbol{Q}}^{0}$ ($\omega_{\boldsymbol{Q}}^{0}$ is
$\omega_{\boldsymbol{Q}}$ at $T=0)$ are $0.86,0.76$ and $0.60$ for $S=1/2,1$
and $\infty,$ correspondingly. Experimentally in the BCC Fe $\omega
_{\boldsymbol{Q}}(T_{C})/$ $\omega_{\boldsymbol{Q}}(0.3T_{c})\approx0.84$
\cite{lynn} and the difference between $\omega_{\boldsymbol{Q}}^{0}$ and
$\omega_{\boldsymbol{Q}}(0.3T_{c})$ is about $15\%$ \cite{lynn}, so the
overall SW renormalisation factor becomes $0.71,$ and our result for $S=1$ is
close to that. Fig. 2 also indicates that $\omega_{\boldsymbol{q}}$ for
smaller spins is less affected at elevated temperatures, implying that QSE
favors the persistence of SW modes.

Let us now estimate the influence of dynamic effects and obtain the relaxation
function $F(\boldsymbol{q},\omega)$. Among various analytical approximations
for $F(\boldsymbol{q},\omega)$ the three-pole approximation \cite{lovesey}
seems to be one of the best and it has been successfully applied to the
typical Heisenberg system with large spin $S=7/2$ \cite{young,latacz}. In this
approximation $F(\boldsymbol{q},\omega)$ is expressed in terms of $\delta
_{1}^{\boldsymbol{q}}=\langle\omega^{2}\rangle_{\boldsymbol{q}}$ and
$\delta_{2}^{\boldsymbol{q}}=\langle\omega^{4}\rangle_{\boldsymbol{q}}%
/\delta_{1}^{\boldsymbol{q}}-$ $\delta_{1}^{\boldsymbol{q}}$, where
$\langle\omega^{n}\rangle_{\boldsymbol{q}}$ are frequency moments of $F$
depending on the static correlation. The evaluation of $\langle\omega
^{2}\rangle_{\boldsymbol{q}}$ is straightforward \cite{lovesey}.
$\langle\omega^{4}\rangle_{\boldsymbol{q}}\propto\langle{\Large [}\ddot
{S}_{\boldsymbol{q}}^{z},i\dot{S}_{-\boldsymbol{q}}^{z}{\Large ]}\rangle$
\cite{loveluck} contains four-spin correlation terms which have to be properly
decoupled as a product of two-spin correlations. In the literature the
conventional decoupling $\langle S_{i}^{+}S_{l}^{z}S_{m}^{z}S_{j}^{-}%
\rangle\rightarrow C_{ij}C_{lm}^{zz}$ $\ $and $\langle S_{i}^{+}S_{l}^{+}%
S_{m}^{-}S_{j}^{-}\rangle\rightarrow C_{im}C_{lj}+C_{ij}C_{lm},$ appropriate
for large $S$, have been applied to obtain $\langle\omega^{4}\rangle
_{\boldsymbol{q}}$ \cite{lovesey,young}. \ For small $S$, the spin kernel
effect, which is neglected in this decoupling, becomes important. This QSE can
be clearly seen in $S=1/2$ case, where for $i=l$ or $m=j$ \ the left side of
the decoupled equation vanishes while the right side is finite. To take into
account this QSE we introduce the following decoupling procedure%
\begin{gather}
\langle\{S_{i}^{+},S_{l}^{z}\}\{S_{m}^{z},S_{j}^{-}\}\rangle\rightarrow
f_{il}^{s}f_{mj}^{s}C_{ij}C_{lm}^{zz}\text{ \ for }R_{il}\leq R_{im}%
,R_{lj},\nonumber\\
\langle S_{i}^{+}S_{l}^{+}S_{m}^{-}S_{j}^{-}\rangle\rightarrow f_{il}%
^{s}f_{mj}^{s}{\large [}f_{ij}^{s}f_{lm}^{s}C_{im}C_{lj}+f_{im}^{s}f_{lj}%
^{s}C_{ij}C_{lm}\nonumber\\
+(\delta_{ij}\delta_{lm}+\delta_{im}\delta_{lj})C_{il}^{zz}%
],\label{decouple41}%
\end{gather}
where $f_{il}^{s}=1-\delta_{il}/2S.$ If $i,l,m$ and $j$ are four different
sites then Eq.(\ref{decouple41}) is the same as in the conventional
decoupling. QSE occurs when two or more out of these four sites are the same.
In this case Eq.(\ref{decouple41}) at $S=1/2$ is exact and is reduced to the
conventional decoupling for $S\rightarrow\infty$. With these results for two
opposite limits of $S$\ and the introduced earlier quantum correction in
$f_{ii}^{s}$ $\sim1/S,$ one can expect that Eq.(\ref{decouple41}) will be a
reasonable interpolation for arbitrary $S$. By applying this decoupling
procedure one can obtain $\langle\omega^{4}\rangle_{\boldsymbol{q}}%
=\langle\omega^{4}\rangle_{\boldsymbol{q}}^{(0)}+\langle\omega^{4}%
\rangle_{\boldsymbol{q}}^{(1)},$ where $\langle\omega^{4}\rangle
_{\boldsymbol{q}}^{(0)}$ corresponds to the conventional decoupling
\cite{lovesey,young} while $\langle\omega^{4}\rangle_{\boldsymbol{q}}^{(1)}$
is the quantum correction given by%

\begin{gather}
\langle\omega^{4}\rangle_{\boldsymbol{q}}^{(1)}=\frac{1}{4S\chi
_{\boldsymbol{q}}}{\LARGE \{}\frac{1}{N}\sum_{\boldsymbol{k}}{\Large [}%
J^{\boldsymbol{k}}(4g_{\boldsymbol{k}}^{2}-6g_{\boldsymbol{k}}%
g_{\boldsymbol{q+k}}+2g_{\boldsymbol{q+k}}^{2})\nonumber\\
-C^{\boldsymbol{k}}(h_{\boldsymbol{k}}-h_{\boldsymbol{k}+\boldsymbol{q}%
})(13J^{\boldsymbol{k}}-7J^{\boldsymbol{q+k}}){\Large ]}-(11g_{\boldsymbol{0}%
}-9g_{\boldsymbol{q}})(h_{\boldsymbol{0}}-h_{\boldsymbol{q}})\nonumber\\
+S^{-1}\sum_{n}z_{n}J_{n}^{3}C_{n}(1-\gamma_{n}^{\boldsymbol{q}}%
)(5C_{n}+7C_{0}-6S){\LARGE \}},\label{w4k1}%
\end{gather}
where $\chi_{\boldsymbol{q}}$ is the $q$-dependent susceptibility,
$g_{\boldsymbol{k}}=\sum_{n}z_{n}J_{n}C_{n}\gamma_{n}^{\boldsymbol{k}}$, and
$h_{\boldsymbol{k}}=\sum_{n}z_{n}J_{n}^{2}C_{n}\gamma_{n}^{\boldsymbol{k}}$.

As a function of $\omega$ the relaxation function $F(\boldsymbol{q},\omega)$
has either one maximum at $\omega=0,$ if $\delta_{2}^{\boldsymbol{q}}%
>2\delta_{1}^{\boldsymbol{q}},$ or three maxima at $\omega=0$ and $\omega
=\pm\omega_{\boldsymbol{q}}^{\max},$ if $\delta_{2}^{\boldsymbol{q}}%
<2\delta_{1}^{\boldsymbol{q}}$. The latter case is often referred as the SW
peak at $\omega_{\boldsymbol{q}}^{\max}$\cite{shastry,young,latacz}. With such
a definition the criteria of the SW existence for given $\boldsymbol{q}$ is
$\delta_{2}^{\boldsymbol{q}}/\delta_{1}^{\boldsymbol{q}}<2$. Usually
$\omega_{\boldsymbol{q}}^{\max}$ is slightly larger than $\omega
_{\boldsymbol{q}}.$ In the literature the SW peak was also defined as
$\langle\omega\rangle_{\boldsymbol{q}}$ \cite{makivic} which is slightly
smaller than $\omega_{\boldsymbol{q}}$. \ Near the critical value $\delta
_{2}^{\boldsymbol{q}}/\delta_{1}^{\boldsymbol{q}}\lesssim2,$ the maximum of
$F$ at $\omega_{\boldsymbol{q}}^{\max}$ is broad. When $\delta_{2}%
^{\boldsymbol{q}}/\delta_{1}^{\boldsymbol{q}}$ is decreased, the SW peak is
more pronounced. \ In Fig. 3 we plot the magnitude of $\delta_{2}%
^{\boldsymbol{q}}/\delta_{1}^{\boldsymbol{q}}$ for different $S$ as a function
of $\boldsymbol{q}$ at $T_{C}$. At fixed $\boldsymbol{q}$, $\delta
_{2}^{\boldsymbol{q}}/\delta_{1}^{\boldsymbol{q}}$ is always decreased if $S$
becomes smaller. Along the $(qq0)$ direction, the critical values of $q,$ when
$\delta_{2}^{\boldsymbol{q}}/\delta_{1}^{\boldsymbol{q}}=2$ , are
$q_{\text{cr}}\approx0.30\pi,0.51\pi$ and $0.61\pi$ for $S=1/2,1$ and $\infty
$,\ correspondingly. Our value of $q_{\text{cr}}$ for $S=1$ agrees with the
experiment result in BCC Fe, where SW modes above $T_{C}$ exist only above
$q\sim\pi/2$ in (110) direction (Fig. 2 of\ Ref. \cite{lynn}). The SW peaks
were also obtained in the SM\cite{shastry} (with spin independent $\delta
_{2}^{\boldsymbol{q}}/\delta_{1}^{\boldsymbol{q}}$), but the value of
$q_{\text{cr}}$ there is considerably higher. Our calculations indicate that
this theory, which correspond to $S=\infty$, will be applicable if QSE is
properly taken into account. \ At $\boldsymbol{q}=(\frac{\pi}{2}\frac{\pi}%
{2}0)$ for $S=1/2,$ $1$ and $\infty$ the ratio $\delta_{2}^{\boldsymbol{q}%
}/\delta_{1}^{\boldsymbol{q}}$ is approximately $0.93,$ $2.2$ and $3.6,$ which
are respectively well below, close to, and well above the critical value
$\delta_{2}^{\boldsymbol{q}}/\delta_{1}^{\boldsymbol{q}}=2$. The corresponding
dynamic structure factor $S(\boldsymbol{q},\omega)=\omega(1-e^{-\omega
/T})^{-1}\chi_{\boldsymbol{q}}F(\boldsymbol{q},\omega)$ as a function of
$\omega$ is shown in the inset of Fig.3. It is clear that at this
$\boldsymbol{q}$ the well-defined SW exist in the case of $S=1/2,$ the
tendency of SW appears for $S=1,$ and\ there is no SW signal at all for
$S=\infty$. Fig.3 shows that QSE \ favors the persistence of SW with
increasing impact for smaller spins. In many real magnets $S$ is not large (
$S\approx1$ in BCC Fe and $S\approx0.3$ in FCC Ni) and we believe that QSE
plays an important role in the MSRO and the magnetic excitations above $T_{C}%
$, especially in the itinerant magnets.

In conclusion, we analytically demonstrated the presence of MSRO in the
Heisenberg model and identified the importance of quantum spin effect on MSRO
for ferromagnets above $T_{C}$. By extending the second-order Green's function
technique to arbitrary $S$ we found that for a system of small spins the
quantum effects greatly contribute to the MSRO and enhance its influence. The
spin dynamics investigation developed from the conventional method of moments
further confirms that QSE favors the persistence of spin wave excitations. We
demonstrated that this previously neglected QSE removes the long-standing
controversy between theory and experiment regarding the presence of MSRO and
SW in Fe and Ni above $T_{C}$ and clearly indicates that the current
prevailing point of view of finite temperature magnetism should be
reconsidered to properly include MSRO and quantum effects.

\bigskip\newpage

Fig.1. $T_{C}/T_{C}^{\text{MF}}$ and $E_{C}/E_{0}$ as a function of $S$ from
SM (lines), the HTE methods (close symbols) and our formalism (open symbols)
\ in BCC structure in the NN coupling case.

\bigskip

\ Fig.2. The calculated SW spectrum $\omega_{\mathbf{q}}$ for the different
$S$ at $T_{C}.$ The dashed line is $\omega_{\mathbf{q}}$ at $T=0$ (FM case).
The inset shows $\cos\theta_{n}$ from nearest to fifth-nearest neighbors.

\bigskip

Fig.3. The calculated $\delta_{2}^{\boldsymbol{q}}/\delta_{1}^{\boldsymbol{q}%
}$ for the different $S$ at $T_{C}$ as a function of $\mathbf{q}$. The
criteria of SW $\delta_{2}^{\boldsymbol{q}}/\delta_{1}^{\boldsymbol{q}}=2$ is
marked by the dashed line.The inset shows $S(\mathbf{q},\omega)$ at
$\mathbf{q=(}\frac{\pi}{2}\frac{\pi}{2}0).$

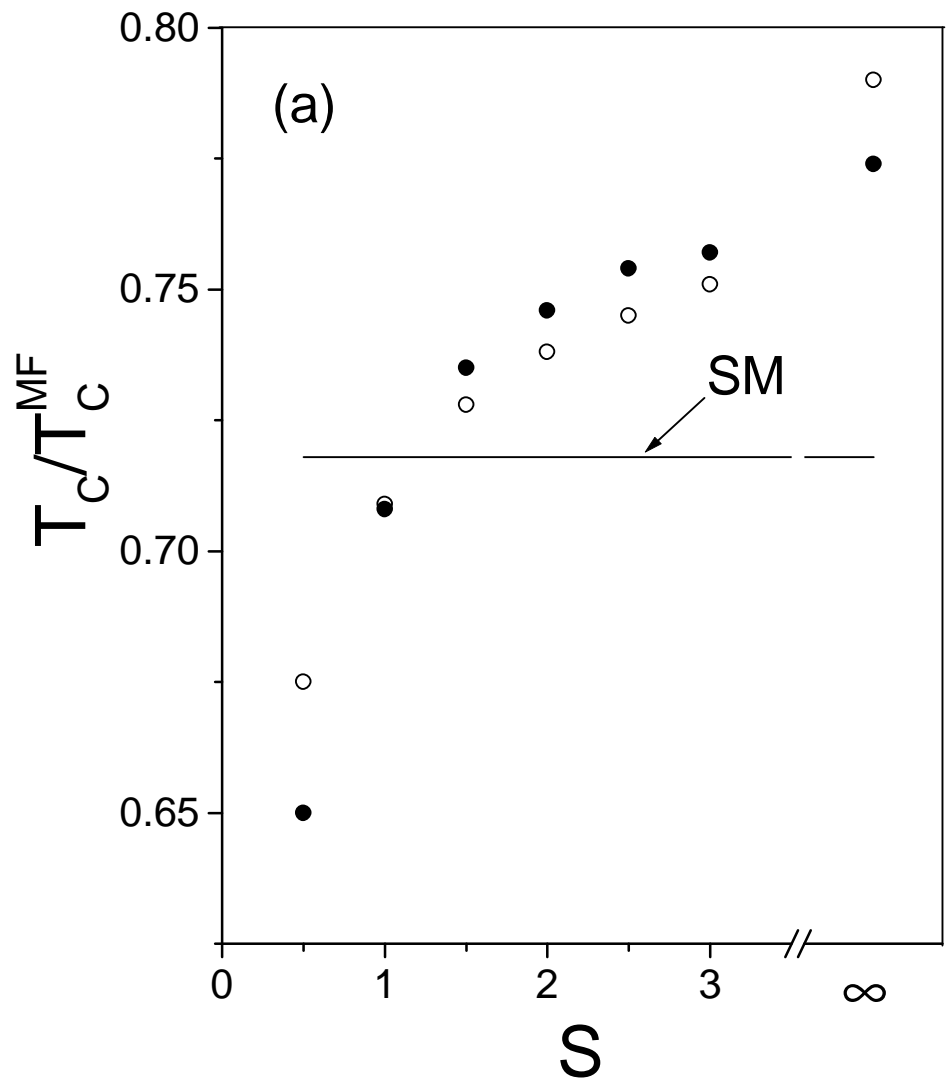
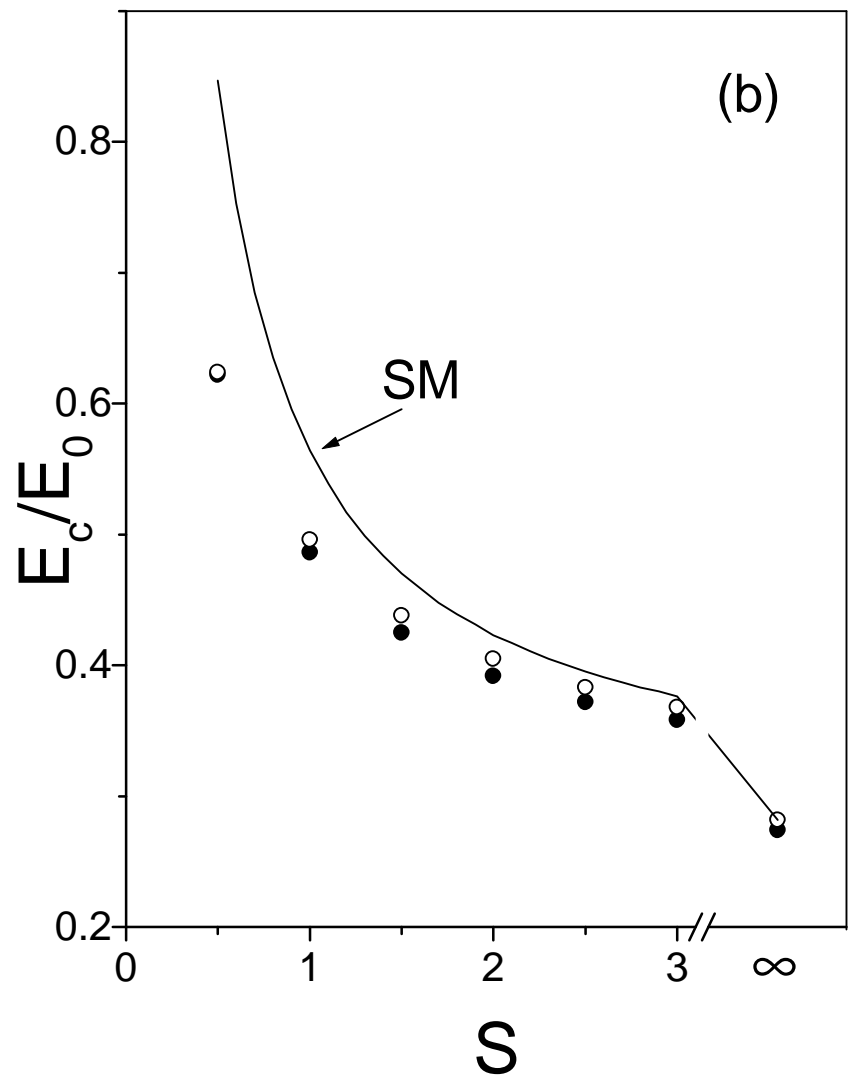
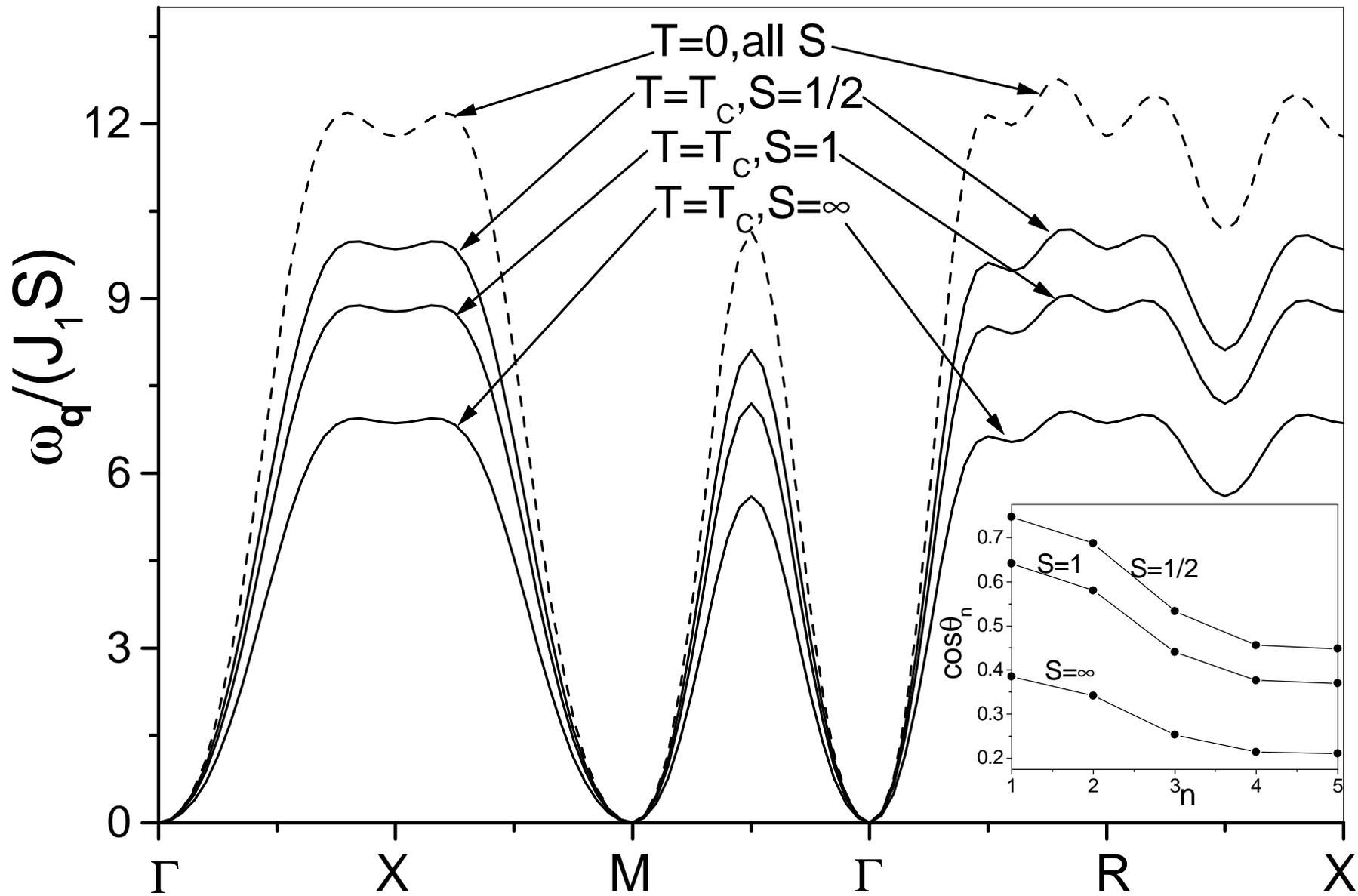
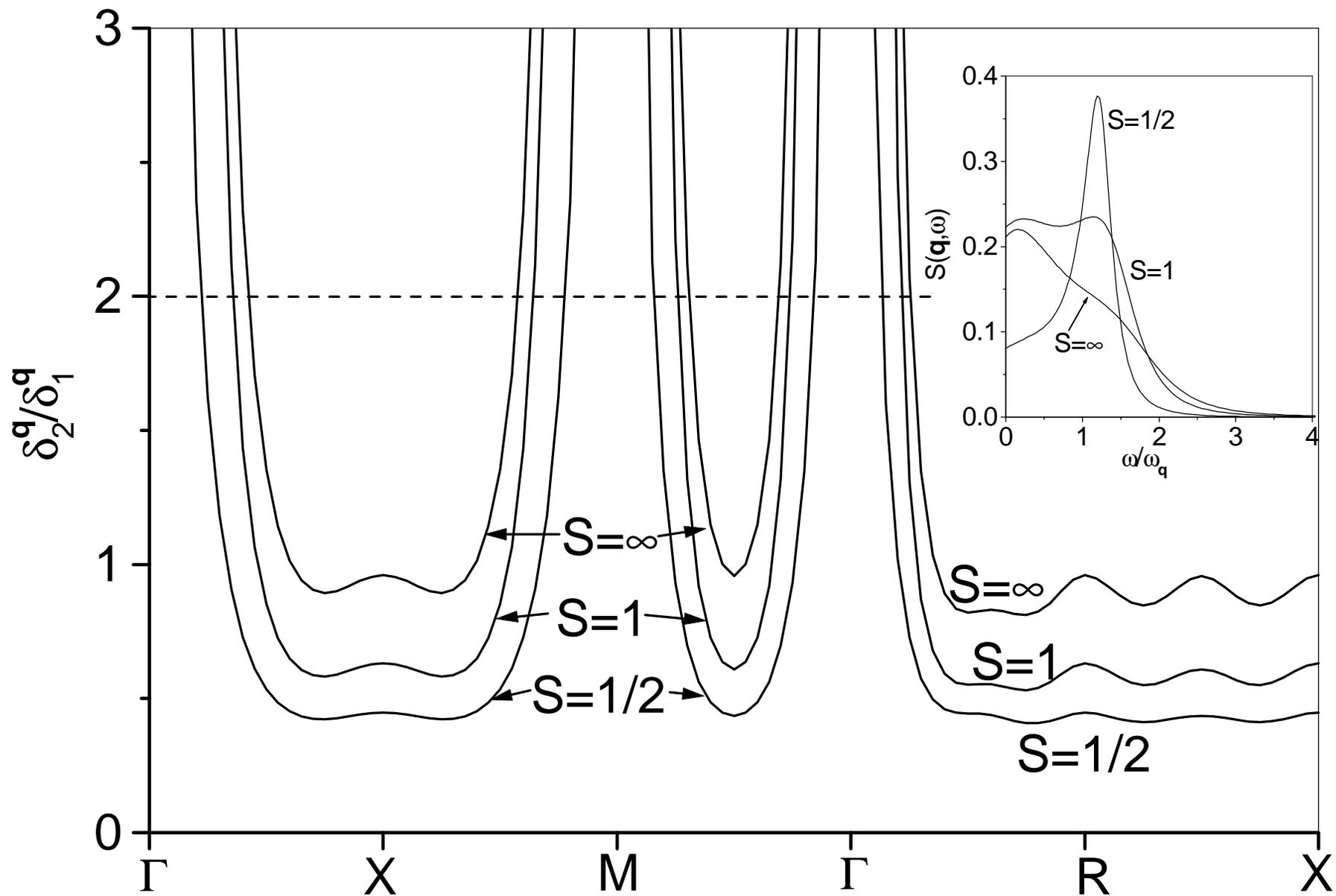
\end{document}